\pdfoutput=1
\documentclass[11pt]{article}

\usepackage[final]{acl}

\usepackage{times}
\usepackage{latexsym}

\usepackage[T1]{fontenc}

\usepackage[utf8]{inputenc}

\usepackage{microtype}

\usepackage{inconsolata}

\usepackage{graphicx}

\usepackage{multirow}
\usepackage{booktabs}
\usepackage{listings}
\usepackage{bbding}
\usepackage{array}
\usepackage[most]{tcolorbox}
\usepackage{enumitem}
\setenumerate[1]{itemsep=0pt,partopsep=0pt,parsep=\parskip,topsep=5pt}
\setitemize[1]{itemsep=0pt,partopsep=0pt,parsep=\parskip,topsep=5pt}
\setdescription{itemsep=0pt,partopsep=0pt,parsep=\parskip,topsep=5pt}

\title{ProjectEval: A Benchmark for Programming Agents Automated Evaluation on Project-Level Code Generation}

\author{
    \textbf{Kaiyuan Liu\textsuperscript{1,2}},
    \textbf{Youcheng Pan\textsuperscript{2}}\Envelope,
    \textbf{Yang Xiang\textsuperscript{2}},
    \textbf{Daojing He\textsuperscript{1}}\Envelope,
    \textbf{Jing Li\textsuperscript{1}}
    \\
    \textbf{Yexing Du\textsuperscript{1,2}},
    \textbf{Tianrun Gao\textsuperscript{3}},
\\
    \textsuperscript{1}Harbin Institute of Technology, Shenzhen, China\\
    \textsuperscript{2}Pengcheng Laboratory, Shenzhen, China  \\
    \textsuperscript{3}Harbin Institute of Technology, Harbin, China\\
   \small{ 
   \href{mailto:1171000408@stu.hit.edu.cn}{1171000408@stu.hit.edu.cn},
   \href{mailto:daojinghehit@163.com}{daojinghehit@163.com},
   \href{mailto:panych@pcl.ac.cn}{panych@pcl.ac.cn},
   }
}

\begin{document}
\maketitle
\begin{abstract}
Recently, LLM agents have made rapid progress in improving their programming capabilities. However, existing benchmarks lack the ability to automatically evaluate from users' perspective, and also lack the explainability of the results of LLM agents' code generation capabilities. Thus, we introduce ProjectEval, a new benchmark for LLM agents project-level code generation's automated evaluation by simulating user interaction. ProjectEval is constructed by LLM with human reviewing. It has three different level inputs of natural languages or code skeletons. ProjectEval can evaluate the generated projects by user interaction simulation for execution, and by code similarity through existing objective indicators. Through ProjectEval, we find that systematic engineering project code, overall understanding of the project and comprehensive analysis capability are the keys for LLM agents to achieve practical projects. Our findings and benchmark provide valuable insights for developing more effective programming agents that can be deployed in future real-world production.\footnotemark[1]
\renewcommand{\thefootnote}{}
\footnotetext{\Envelope ~Corresponding authors}
\footnotetext{$^1$Dataset, code and constructed evaluation machine are available at \href{https://github.com/RyanLoil/ProjectEval/}{https://github.com/RyanLoil/ProjectEval/}.}
\end{abstract}

\section{Introduction}
\begin{table*}
    \centering
    \renewcommand{\arraystretch}{1.5}
    \setlength{\tabcolsep}{3pt}
    \renewcommand\arraystretch{1}
    \small
    \resizebox{\linewidth}{!}{
    \begin{tabular}{ccccccccccccc}
    \toprule
        \multirow{2}{*}{\textbf{Benchmarks}} & \multirow{2}{*}{\textbf{Subject}} & \multirow{2}{*}{\textbf{Test Type}} &\textbf{Construction}&\multicolumn{2}{c}{\textbf{Input Level}}&\multicolumn{2}{c}{\textbf{Evaluation}}& \textbf{Pass@k}& \textbf{\#Tasks}&\textbf{\#Tests}&\textbf{\#LOC}&\textbf{\#Tokens}\\
         & & & Automated & Checklist & Skeleton & Automated & LLM-less  & Available  \\ 
        \midrule
        SoftwareDev\citeyearpar{hong2024metagpt} & Execution& - & \XSolidBrush & \XSolidBrush & \XSolidBrush & - & -  & \XSolidBrush & 70&-&191.6&6218.0\\ 
        ProjectDev\citeyearpar{nguyen2024agilecoder} & Execution & Manual Checklist Reviewing &\XSolidBrush & \Checkmark & \XSolidBrush & \XSolidBrush & \Checkmark  & \Checkmark & 14&19.10&-&36818.0\\ 
        SRDD\citeyearpar{qian2024srdd} & Checklist & LLM-Rating & \Checkmark & \Checkmark & \XSolidBrush & \Checkmark & \XSolidBrush  & \XSolidBrush & 1200&-&-&-\\ 
        CASSD\citeyearpar{zhang2024cassd} & Execution & Manual Checklist Reviewing & \XSolidBrush & \Checkmark & \XSolidBrush & \XSolidBrush & \Checkmark & \Checkmark & 72 & 5.25 & $\approx$240 & 21993.0\\
        \midrule
        DevBench\citeyearpar{li2024devbench} & Execution & Inside Project Test Units & \XSolidBrush & \XSolidBrush & \XSolidBrush& \Checkmark & \Checkmark & \Checkmark & \textbf{22}&10.18&377.8&1298.3\\
        \textbf{ProjectEval(ours)} & Execution & \textbf{User Interaction Simulation} & \CircleSolid & \Checkmark & \Checkmark & \Checkmark & \Checkmark  & \Checkmark & 20&\textbf{14.2}&\textbf{402.20}&\textbf{2972.0}\\ 
        \bottomrule
    \end{tabular}
    }
    \caption{Summary of Existing Project Level Benchmarks. \#Tasks: number of tasks, \#Tests: average number of testcase in each task, \#LOC: average lines of code in the canonical answer, \#Tokens: average number of tokens of code in the canonical answer. \scalebox{0.75}{\CircleSolid}: Our benchmark is constructed by GPT-4o with human reviewing and editing, which is semi-automated. -:SoftwareDev does not contain any evaluation. }
    \label{tab:compare}
\end{table*}
The field of programming has seen significant advances with the rise of large language models (LLMs), today the LLM agents can do many programming works without the help of humans \citep{liu2024exploring}. To evaluate the programming ability of GPT-2, \citet{chen2021humaneval} raised the first programming benchmark named HumanEval in 2021. HumanEval was also the first of the HumanEval-based benchmarks. 
In the last three years, more than 20 benchmarks had been raised manually or automatically based on LLM. 
After HumanEval, MBPP \citep{austin2021mbpp} came out which also was a base for many benchmarks. MBPP concentrating on algorithm realization.  
In 2023, DS-1000 \citep{lai2023ds1000} was raised and represented a new series of benchmarks, evaluating LLM agents programming abilities of accessing the third-party libraries or packages.
In 2024, LLM-based programming agents developed rapidly, researchers noticed that many of them can do project-level programming (\citealp{hong2024metagpt}, \citealp{nguyen2024agilecoder}). 
Therefore, project-level benchmarks came out, they were SoftwareDev \citep{hong2024metagpt}, ProjectDev \citep{nguyen2024agilecoder}, SRDD \citep{qian2024srdd}, CASSD \citep{zhang2024cassd}, and DevBench \citep{li2024devbench}. These benchmarks provided the methods evaluated agent's code generation capabilities at the granularity level project. \textbf{However, only DevBench achieved automated evaluation using in-project test units, and the test-unit-evaluation was NOT a realistic method in a production environment used by human users. And, the other benchmarks still relied on natural language tests, which usually required human judgment for correctness and efficiency. Moreover, only DevBench provided a table of evaluation scores, but it lacked explainability.}

To bridge this gap, we propose \textbf{ProjectEval}, a novel benchmark tailored for automated evaluating project-level programming tasks(missions) in this field. ProjectEval is designed to assess the ability of agents to tackle complex user-driven tasks with precision and adaptability. Unlike existing benchmarks, ProjectEval emphasizes real-world usability by \textbf{integrating automated test suites of \underline{user interaction simulation}, parameter analysis, and canonical solutions} into a cohesive evaluation pipeline. ProjectEval contains 20 real-world tasks with totally 284 testcases, and supports two task types: website-based projects and batch/console-based programs, with theoretical scalability to even more complex, custom UI-based tasks. By leveraging LLMs for generation, supplemented with manual reviewing, ProjectEval ensures robust and detailed evaluation metrics.

We introduce the core structure and evaluation process of ProjectEval, highlight the challenges faced by state-of-the-art models such as Gemma-2 and GPT-4o, and demonstrate the benchmark's ability to test comprehensive capabilities through pass rate (Pass@K) and existing algorithms. Our goal is to establish a more effectively, and more explainable new standard for automated evaluating project code quality.
We summarize the contributions of this work as follow:
\begin{itemize}
    \item User-Centric Project-level Benchmark: ProjectEval fills the gap in existing benchmarks by offering a user-focused framework with real-world applicability with comprehensive and project-level metrics. It supports website and batch/console-based projects.
    \item Enhancing the Agent Code Generation Explainability by Three Different Level Inputs: Three different level inputs integrating three kinds of objective indicators and pass rate, ProjectEval ensures precise, adaptable evaluation and enhances the result explainability. 
    \item Automated Evaluation Testsuites: ProjectEval realizes the automated evaluation from user perspective in website tasks and batch/console tasks through simulating user interaction. This is a new low-cost method to evaluate the code generation capabilities of agents. 
\end{itemize}

\section{Related Works}
\subsection{LLM-Based Coding Agents}
There are many code LLMs to date, e.g. StarCoder, InCoder, WizardCoder, CodeGen and etc. However, only the NL-i/o-available models could become programming agents as the instruction of an agent-level mission used to be natural language. Without NL input ability, the LLMs are really hard to become agents through agent designs of CoT, ReAct or Reflextion. Some researchers have developed LLM agents. ChatDev \citep{qian2024srdd} presents a diffusion-based model combining large language models with image decoders, advancing text-to-image generation. AgileCoder \citep{nguyen2024agilecoder} introduces multi-agent Agile role assignments for efficient, collaborative software development, while MetaGPT \citep{hong2024metagpt} encodes workflows into prompt sequences for structured multi-agent task management. Besides some large multimodal reasoning models also have the abilities to generate codes \citep{li2025multimodal}.
\subsection{Benchmarks for Code Generation}
\begin{table*}[ht!]
\centering
\setlength{\tabcolsep}{1pt}
\renewcommand\arraystretch{0.95}
\scriptsize
\begin{tabular}{@{}c|ccccm{1.4cm}<{\centering}ccccc@{}}
\toprule
\textbf{Benchmark} & \textbf{Language} & \textbf{Construction} & \textbf{Evaluation} & \textbf{Source} & \textbf{Granularity Level} & \textbf{\#Tasks} & \textbf{\#Tests} & \textbf{\#LOC} & \textbf{\#Tokens} & \textbf{Input Information} \\
\midrule
\multicolumn{11}{c}{HumanEval-based}\\
\midrule
HumanEval \citeyearpar{chen2021humaneval} & Python & Manual & Automated & Original & Function & 164 & 7.7 & 11.5 & 24.4 & NL+ Signature \\
Multi-HumanEval\citeyearpar{athiwaratkun2023multihumaneval} & Multiple & Manual & Automated & HumanEval \& Original & Function & 164 & 7.7 & 11.5 & 24.4 & NL+ Signature \\
\midrule
\multicolumn{11}{c}{MBPP-based}\\
\midrule
MBPP\citeyearpar{austin2021mbpp} & Python & Manual & Automated & Original & Function & 974 & 3.0 & 6.8 & 24.2 & NL \\
CodeContests\citeyearpar{li2022codecontests} & Python, C++ & Automated & Automated & Contest Sites & Competitive & 165 & 203.7 & 59.8 & 184.8 & NL + Example I/O \\
\midrule
\multicolumn{11}{c}{DS-based}\\
\midrule
DS-1000\citeyearpar{lai2023ds1000} & Python & Automated & Automated & Stack Overflow & Statement & 1000 & 1.6 & 3.8 & 12.8 & NL \\
CoderEval\citeyearpar{yu2024codereval} & Python, Java & Automated & Automated & Github & Function & 230 & N/A & 30.0 & 108.2 & NL + Signature \\
SWEBench\citeyearpar{jimenez2024swebench} & Python & Automated & Automated & PyPI & Commit & 2294 & 120.8 & 32.8 & $\approx$200 & Issue + Repository \\
ClassEval \citeyearpar{du2023classeval}& Python & Manual & Automated & PyPI + Original & Class & 100 & 33.1 & 45.7 & 123.7 & Class Skeleton \\
\midrule
\multicolumn{11}{c}{Granularity Level - Project}\\
\midrule
SRDD \citeyearpar{qian2024srdd}& Python & Automated & Automated & Original & Project & 1200 & N/A  & N/A  & N/A  & NL \\
CASSD \citeyearpar{zhang2024cassd} & Python & Manual & Manual& Original & Project & 72 & 5.25 & $\approx$240 & 21993.0 & NL \\
SoftwareDev\citeyearpar{hong2024metagpt} & Multiple & Manual & N/A & Original & Project & 70 & N/A  & 191.6 & 6218.0 & NL \\
ProjectDev\citeyearpar{nguyen2024agilecoder}  & Multiple & Manual & Manual & Original & Project & 14 & 19.1 & N/A  & 36818.0 & NL \\ 
DevBench\citeyearpar{li2024devbench} & Multiple & Manual & Automated & Original & Project & 22 & 10.18 & 377.8 & 1298.3 & NL \\
\multirow{2}{*}{\textbf{ProjectEval (ours)}} & \multirow{2}{*}{\textbf{Python}} & \multirow{2}{*}{\textbf{Semi-automated}} & \multirow{2}{*}{\textbf{Automated}} & \textbf{SoftwareDev \&} &  \multirow{2}{*}{\textbf{Project}} & \multirow{2}{*}{\textbf{20}} & \multirow{2}{*}{\textbf{14.2}} & \multirow{2}{*}{\textbf{402.2}} & \multirow{2}{*}{\textbf{2972.0}} & \textbf{NL+ Class Skeleton} \\
&&&&\textbf{ProjectDev \& Origin}&&&&&&\textbf{/Function Skeleton}\\
\bottomrule
\end{tabular}
\caption{Summary of Existing Benchmarks for Code Generation. \#Tasks: number of tasks, \#Tests: average number of testcase in each task, \#LOC: average lines of code in the canonical answer, \#Tokens: average number of tokens of code in the canonical answer. N/A: This benchmark doesn’t involve this item. Part of this table is referred from \citeposs{du2023classeval}. Other categories are in Appendix~\ref{sec:benchmarks-fullversion}.}
\label{tab:benchmarks}
\end{table*}

Table~\ref{tab:benchmarks} provides a detailed summary of existing benchmarks in the programming domain, and categorize them into six groups: exploratory benchmarks before 2020, HumanEval-based benchmarks (\citealt{chen2021humaneval},\citealt{liu2024is}, \citealt{hao2022aixbench}, \citealt{peng2024humaneval-xl}, \citealt{athiwaratkun2023multihumaneval}), MBPP-based benchmarks (\citealt{austin2021mbpp}, \citealt{peng2024humaneval-xl}, \citealt{hendrycks2021measuring}, \citealt{huang2024effibench}, \citealt{jain2024livecodebench}, \citealt{li2023taco}, \citealt{li2022codecontests}), DS-based benchmarks (\citealt{lai2023ds1000}, \citealt{du2023classeval}, \citealt{yu2024codereval}), problem understanding benchmarks, and those focused on specific domains or methods. With the development of LLM and agents, the benchmarks are gradually moving toward higher granularity level. The project-level is the last and highest level of programming. The project-level benchmarks, which are the primary focus of our research, comprehensively evaluate the process of transforming an initial idea into a complete project. 

There are currently five benchmarks in this category, see Table~\ref{tab:compare} for the differences. In summary, all project-level benchmarks can't evaluate automatically except DevBench and our ProjectEval. Except ours, all benchmarks don't provide a multi-level inputs. Compared with DevBench, we have more testcases, and our tasks (missions) are complicate than DevBench as more lines of code and more tokens. Besides, existing project-level benchmarks use the test units, manual checklist reviewing or LLM-scoring directly from the tested library or framework to complete the evaluation, rather than actually compiling and executing the project and checking in users' perspective.

For other based benchmarks, see Table~\ref{tab:benchmarks} for brief and Appendix~\ref{sec:benchmarks-fullversion} for full version.

\section{ProjectEval Benchmark}
\subsection{Benchmark Format}
\begin{figure}[t]
    \centering
    \includegraphics[width=1\linewidth]{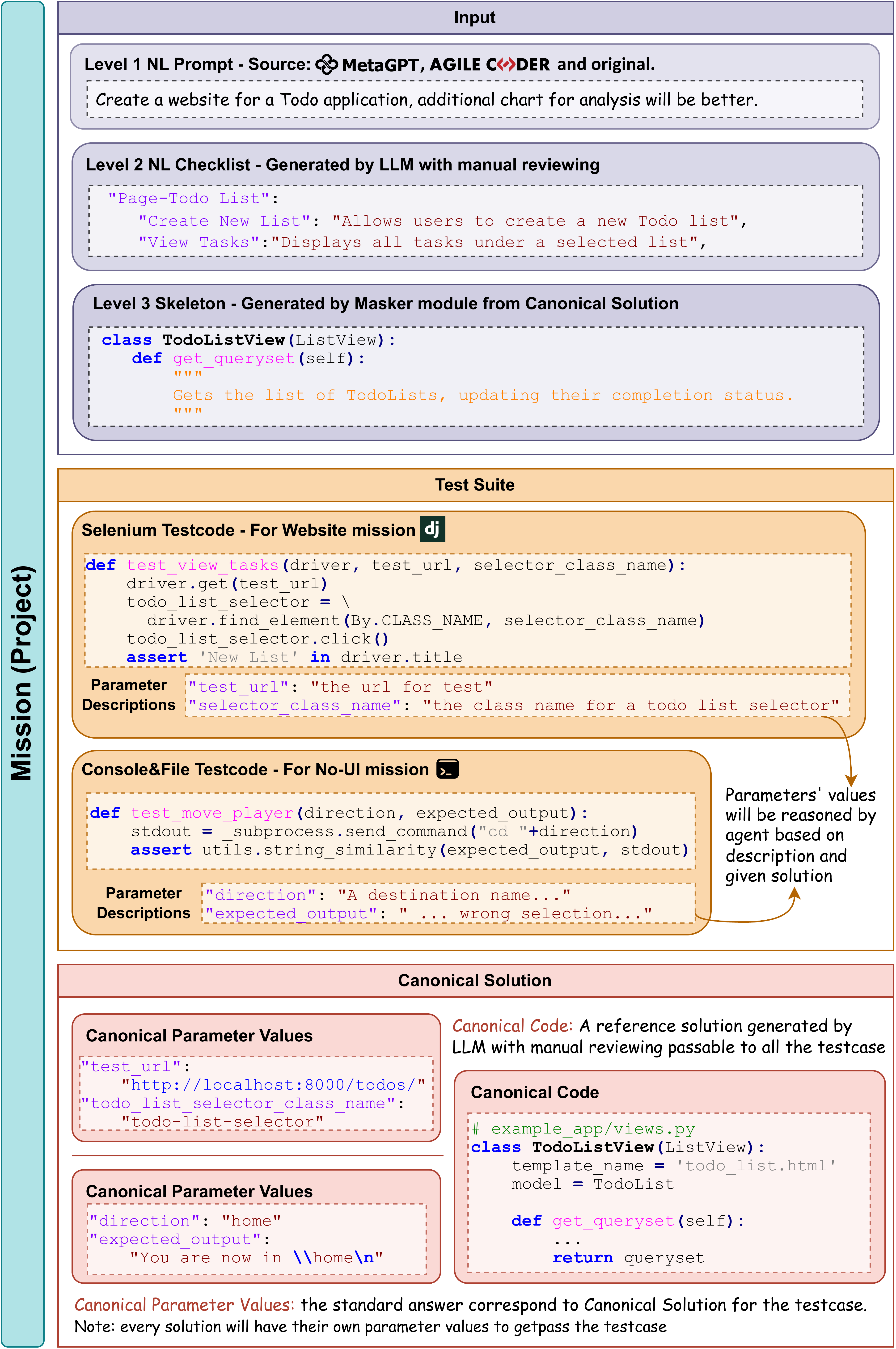}
    \caption{A typical ProjectEval website mission, including three different levels of input, a test suite, and a canonical solution. Notice that the upper test suite is the test suite used in website mission while the lower one is an example of console/file mission test suite.}
    \label{fig:structure}
\end{figure}

A standard ProjectEval mission will have three parts: Inputs, Test Suite and Canonical Solution. Figure~\ref{fig:structure} shows an example structure of a standard ProjectEval mission.

As for the inputs, there are three different input types named Level for the test in each mission for the agent to achieve the target in ProjectEval (See example in Figure~\ref{fig:structure} purple part):
\begin{itemize}
    \item \textbf{Level 1 - Natural Language Prompt (NL Prompt)}: In this level, the agent will receive one or several natural language sentences to describe the target of the project. The agent will create the entire project ONLY based on these sentences.
    \item \textbf{Level 2 – Natural Language Checklist (NL Checklist)}: In this level, the agent will receive a standard natural language checklist describing the project through the abilities and functions that the project should have.
    \item \textbf{Level 3 – Skeleton}: In this level, the agent will receive a skeleton of the standard answer. This skeleton contains doc-strings and comments to describe the project inside.
\end{itemize}
A mission test suite will contain two parts (See example in Figure~\ref{fig:structure} orange part):
\begin{itemize}
    \item \textbf{Testcodes}: a mission contains several automated evaluation Python functions similar to HumanEval testcases. But, these testcodes are prohibited using test unit inside the technical stack but using user simulation by operating UI to test the project generated by agents.
    \item \textbf{Parameter Description (PD)}: usually, every testcode has a matching parameter descriptions. PD is used for a special kind of parameter alignment. These parameters are required by the matching testcode to achieve the established test goal(s), \emph{e.g.} in Figure~\ref{fig:structure}, the “test\_url” is the URL of the page which can show all the “tasks” that are required by the testcodes. PD is similar with a user manual given by developers to guide users to accomplish what they want to do - that is, the main evaluation concept we designed: evaluation based on the user's perspective.
\end{itemize}
Finally, every mission we constructed has a canonical solution, beside the canonical code, we also build every PD’s standard answer matching to the canonical code called canonical parameter values (See example in Figure~\ref{fig:structure} red part).
In addition, we categorized tasks into “easy”, “medium”, “hard”, and “human” based on code volume and human-reviewed complexity. However, since results show no distinction, we won’t elaborate further.

\textbf{We give the agent missions by JSON format directly embedded into their prompt and asked the same format output, which we consider as a very important ability of code generation.}

\subsection{Construction Process}
\label{subsec:construction}
The construction process of ProjectEval is relatively complex (See Appendix~\ref{sec:full-figure} for the complete version ProjectEval process and structure diagram). 

\begin{figure*}[t]
    \centering
    \includegraphics[width=1\linewidth]{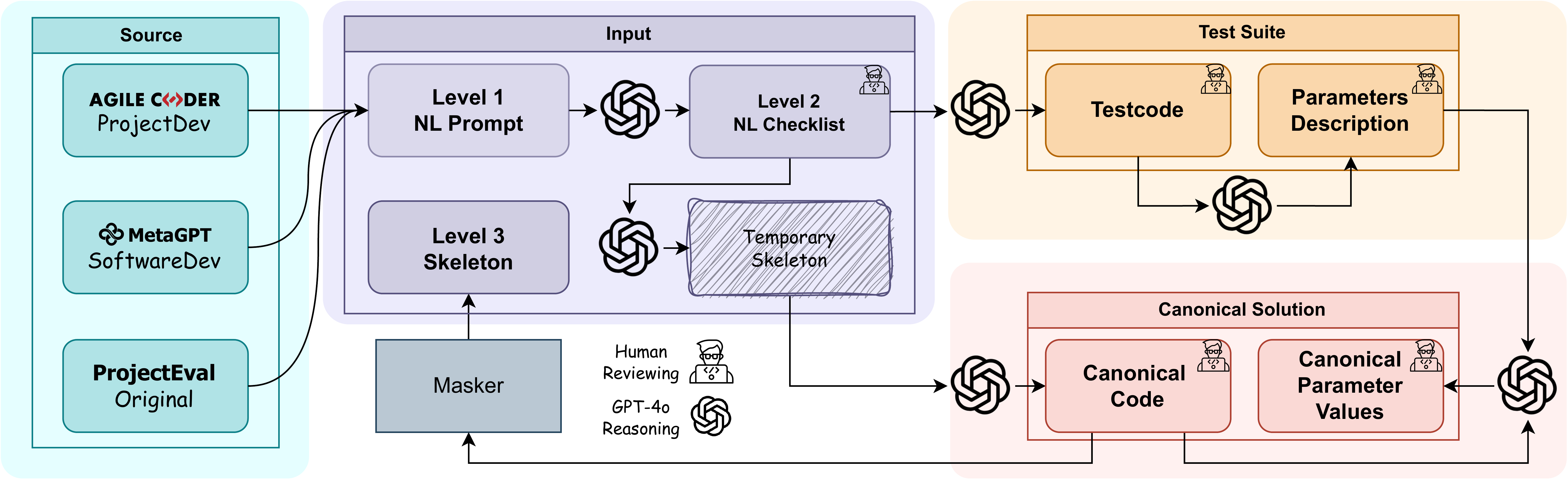}
    \caption{Construction of ProjectEval. 
    Testcode is aligned with Checklist. Parameter Description is aligned with Testcode and Canonical Parameter Values. Canonical Parameter Values is aligned with Canonical Code and use for testcode to get passed.}
    \label{fig:generation}
\end{figure*}

\textbf{Level 1 NL Prompt \& Level 2 NL Checklist:} There are initial 20 tasks (missions) in ProjectEval that are manually edited into concise natural language descriptions, which is Level 1 NL Prompt. 7 of them are sourced from SoftwareDev \citep{hong2024metagpt} and ProjectDev \citep{nguyen2024agilecoder}  while others are created originally by us. Figure~\ref{fig:generation} purple part shows that these descriptions are sent to an LLM, which generates a list of more detailed natural language task descriptions. After manual review and modification, the refined version is referred as the Level 2 NL Checklist. 

\textbf{Testsuite:} Figure~\ref{fig:generation} orange part shows that the NL Checklist is given into the LLM, which, from a user testing perspective, generates test code. \textbf{For website missions, the test code is mostly implemented using the open-source testing library Selenium, which simulates user behavior in a browser to interact with websites. For batch/console tasks, the test code typically uses Python's subprocess module to mimic user interactions such as running commands and entering keyboard input.} If the task involves file generation, the test code utilizes dedicated open-source libraries to read and compare the similarity of the generated file against a canonical file. For example, programs that generate Excel files are validated using the Openpyxl library to compare with the reference files. The test code often requires one or more parameters to execute because the specifics of the code generated by an agent—such as variable names, function names, class names, and output file names—are unpredictable. To address this, the test code is input into the LLM to generate an additional Parameter Description (PD), which provides a natural language explanation of the parameters needed by the test code. The PD, along with the test code, constitutes the Test Suite.

\textbf{Canonical Solution:} Simultaneously, in Figure~\ref{fig:generation} red part, the NL Checklist is put into another LLM thread to generate a temporary project skeleton, which is then fed back into the LLM to infer and generate Canonical Code (CNC). Practical results show that while most of the code can't be use directly, a little of the LLM-generated code is mostly correct, but it often requires manual corrections to form the true canonical code. This process aligns with the findings of AgileCoder experiments partially. A human reviewer is asked for checking the CNC to confirm that it is runnable and meets the requirements of the Checklist. By inputting the PD and CNC into the LLM and applying minimal manual adjustments, Canonical Parameter Values (CPV) are obtained. Together, CPV and CNC are the Canonical Solution (CNS). When the CNS is input into the ProjectEval testing controller, it achieves a perfect score, \emph{i.e.}, Pass@K = 100\%.

\textbf{Level 3 Skeleton:} Finally, the CNC is processed through a Masker (which could be a regex-based program or an LLM) to replace function bodies, class bodies, and critical HTML tag content with functional description comments. This produces a test skeleton that can evaluate LLMs without natural language generation capabilities, referred to as the Level 3 Skeleton.

All CNC are programmed in Python but \textbf{Project-Eval theoretically supports any programming language} as we evaluate the LLM through users' perspective. It may need researchers compile the LLM-generated program in advance. The total cost of construction process with GPT-4o is \$2.95 and the human reviewing cost is \$420 by hiring a third-party company with contract.

\subsection{Evaluation Process}
\label{subsec:evaluation-process}
\begin{figure*}[t]
    \centering
    \includegraphics[width=1\linewidth]{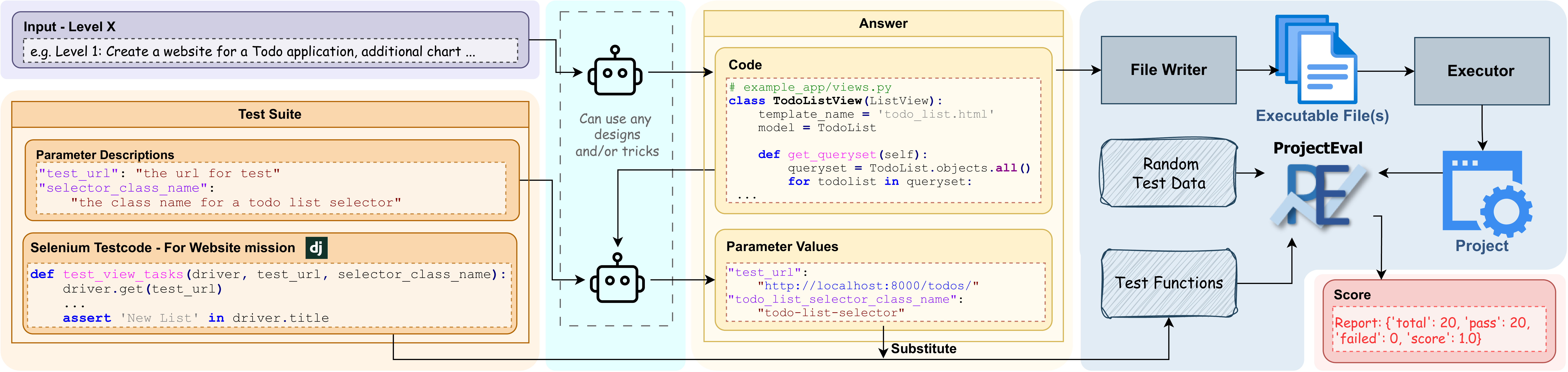}
    \caption{Evaluation Process of ProjectEval. 
    The evaluation process begins by selecting a specific level from the input and presenting it to the agent. The agent generates solution code. The solution code is then fed back into the same agent along with the parameter description. The agent is tasked with answering the parameter description based on its own solution to produce parameter values (PV). The code is then converted into an executable file, creating a tangible project. PV is substitute to testcode, and testcode is integrated into the ProjectEval evaluation machine to obtain the evaluation results.}
    \label{fig:reasoning}
\end{figure*}

The evaluation process begins by selecting a specific level from the input and presenting it to the agent (See Figure~\ref{fig:reasoning}). The agent can use any designs or methods to solve the inputs and generate Solution Code (Code). The Code is then fed back into the same agent along with the PD. The agent is tasked with answering the parameter description based on its own generated Code to produce Parameter Values (PV). The Code is then converted into an executable file, creating a tangible project in a sandbox environment. This project, together with the testcode with PV substituted, is integrated into the ProjectEval evaluation machine to obtain the evaluation results which is also done in a sandbox.
\begin{table}[!htbp]
    \centering
    \renewcommand{\arraystretch}{1.5}
    \small
    \setlength{\tabcolsep}{2pt}
    \renewcommand\arraystretch{1.5}
    \resizebox{\linewidth}{!}{
    \begin{tabular}{lcc}
    \toprule
        \textbf{Parts} & \textbf{Method} & \textbf{Type} \\ \midrule
        \multirow{2}{*}{Level 2 Checklist} & Sentence Transformer \citeyearpar{gurevych2020sentence-transformer} & Maximum  \\
        & + Jonker Volgenant \citeyearpar{jonker1987shortest}&\\
        \multirow{2}{*}{Level 3 Skeleton} & CodeBLEU \citeyearpar{ren2020codebleu}& Maximum \\
        & + Jonker Volgenant & \\
        Code & CodeBLEU + Jonker Volgenant & Maximum \\
        Parameters Values & Levenshtein Distance \citeyearpar{levenshtein1966binary} & Average \\ 
        \bottomrule
    \end{tabular}
    }
    \caption{ProjectEval Objective Indicators. Four additional objective similarity evaluation methods to evaluate the performance of each parts individually.}
    \label{tab:objective-indicators}
\end{table}

Since we have different level inputs, we can compare the similarity of the generated results at each level and obtain the score for each step. This process is equivalent to disassembling the CoT of LLM agents to a certain extent, thereby enhancing the explanabilities of the pass rate results.

Therefore, we introduces four additional objective similarity evaluation methods to evaluate the performance of four parts individually (See Table~\ref{tab:objective-indicators}). As a Level 2 Checklist consists of multiple independent natural language sentences, which cannot be considered a cohesive document, after calculating the similarity between each sentence in the canonical Checklist and the test Checklist using Sentence Transformers \citep{gurevych2020sentence-transformer}, the Jonker-Volgenant algorithm \citeyearpar{jonker1987shortest} is employed to determine the optimal matching scheme, from which an overall matching score is derived. Since both the Level 3 Skeleton and the answer are written as code, existing code evaluation tools like CodeBLEU \citep{ren2020codebleu} are used to compute BLEU scores by considering structure similarity. The Skeleton and Code have the rectangular linear sum assignment problems same as Checklist, so Jonker-Volgenant is also used in these parts. Parameter Values are typically short, often consisting of single words, compound words, or simple URLs. Therefore, strings cosine similarity is directly used to measure their similarity.

\section{Experiments}\label{sec:experiments}
\subsection{Research Questions}
Our experiments intend to answer the following research questions:

\begin{itemize}
    \item \textbf{RQ1 (Overall Correctness)}: How do LLM agents perform on ProjectEval benchmark?
    \item \textbf{RQ2 (Cascade Generation \& Direct Generation)}: Do LLM agents performs better when they generated level by level till answer code (\emph{i.e.} cascade) than directly generate?
    \item \textbf{RQ3 (Basic LLM Selection)}: Which basic LLM perform the best in the experiments and where it does better than the others?
    \item \textbf{RQ4 (Step by Step Performance)}: How do LLM agents performs on each part of ProjectEval benchmark?
\end{itemize}
For the basic LLM selection and settings part, see Appendix~\ref{sec:selection} for the details.
\subsection{Evaluation Metric}
Same as many benchmarks of HumanEval-based and MBPP-based, we adapt the pass rate (Pass@K) for every LLM. We have average 14.2, totally 284 testcases (including runnable as a testcase) to evaluate the correctness of Code generated by LLM (See Table~\ref{tab:compare} for all statistics). The percentage of test cases that passed is the final score that an LLM gains from ProjectEval. Notice that some of the testcases are chain-reacted, as if the former one fails, the followings will never get passed.

We also added 4 objective indicators mentioned in Section~\ref{subsec:evaluation-process} for each part which agents generated mentioned in Table~\ref{tab:objective-indicators}. ProjectEval will compute these metrics in parallel with Pass@K.

The total evaluation cost of Pass@5 with GPT-4o is \$28.02, average \$5.60 for each round.
\begin{table}[!ht]
\centering
    \small
    \setlength{\tabcolsep}{1pt}
    \renewcommand\arraystretch{1}
    \resizebox{\linewidth}{!}{
    \begin{tabular}{l|ccc|cccc|c}
        \toprule
        \multirow{2}{*}{\textbf{Model}}&\multicolumn{3}{c|}{ \textbf{Cascade}} & \multicolumn{4}{c|}{ \textbf{Direct}}&  \multirow{2}{*}{ \textbf{All Avg.} }\\
        \cmidrule{2-8}
        &\textbf{Level 1} & \textbf{Level 2}  & \textbf{Avg.} & \textbf{Level 1} & \textbf{Level 2} & \textbf{Level 3} & \textbf{Avg.} &  \\
        \midrule
        \multicolumn{9}{c}{\textbf{Open-source AGI LLMs}}\\
        \midrule
        Llama-2-7B & 0.00 & 0.07 & 0.04 & 0.28 & 0.00 & 0.07 & 0.12 & 0.08 \\ 
        Llama-3.1-7B & 0.28 & 0.28 & 0.28 & 0.14 & 0.28 & 0.42 & 0.28 & 0.28 \\ 
        Llama-3.2-3B & 0.21 & 0.14 & 0.18 & 0.14 & 0.00 & 0.00 & 0.05 & 0.10 \\ 
        Phi-3-14B & 0.00 & 0.00 & 0.00 & 0.00 & 0.00 & 0.00 & 0.00 & 0.00 \\ 
        Phi-4-14B & 0.14 & 0.56 & 0.35 & \textbf{1.76} & \textbf{1.13} & \textbf{2.04} & \textbf{1.64} & \textbf{1.13} \\ 
        Gemma-7B & 0.99 & 1.06 & 1.02 & 0.56 & 0.63 & 0.49 & 0.56 & 0.75 \\ 
        Gemma-2-9B & \textbf{1.69} & \textbf{1.06} & \textbf{1.37} & 1.34 & 0.56 & 0.63 & 0.85 & 1.06 \\ 
        Mistral-7B-v0.3 & 1.48 & 1.06 & 1.27 & 0.92 & 0.99 & 0.56 & 0.82 & 1.00 \\ 
        \midrule
        \multicolumn{9}{c}{\textbf{Code Generation LLMs}}\\
        \midrule
        StarCoder-2-7B & - & - & - & - & - & 0.00 & - & - \\ 
        CodeGemma & - & - & - & - & - & \textbf{1.20} & - & - \\ 
        CodeLlama & - & - & - & - & - & 0.77 & - & - \\ 
        \midrule
        \multicolumn{9}{c}{\textbf{Close-source AGI LLMs}}\\
        \midrule
        GPT-3.5-turbo & 2.39 & 2.46 & 2.43 & 1.97 & 2.39 & 5.28 & 3.22 & 2.90 \\ 
        GPT-4o & \textbf{\underline{8.52}} & \textbf{\underline{12.32}} & \textbf{\underline{10.42}} & \textbf{\underline{16.06}} & \textbf{\underline{15.42}} & \textbf{\underline{10.14}} & \textbf{\underline{13.87}} & \textbf{\underline{12.49}} \\ 
        Gemini 1.5 pro  & 7.82 & 7.39 & 7.61 & 5.28 & 4.51 & 8.24 & 6.01 & 6.65 \\ 
        Gemini 2.0-flash & 3.24 & 3.59 & 3.42 & 3.52 & 3.45 & 7.75 & 4.91 & 4.31 \\ 
        \midrule
        Avg. & 2.23 & 2.50 & 2.37 & 2.66 & 2.45 & 2.51 & 2.69 & 2.56 \\ 
        \toprule
    \end{tabular}
    }
    \caption{ProjectEval Result Pass@5. ProjectEval is hard for recent LLM agents to get pass. GPT-4o has the best score. See Appendix \ref{sec:pass@k} for Pass@K.}    
    \label{tab:pass@5}
\end{table}
\section{Results}\label{sec:results}
\begin{table*}[ht]
    \centering
    \small
    \renewcommand\arraystretch{1}
    \setlength{\tabcolsep}{4pt}
    \resizebox{\textwidth}{!}{
        \begin{tabular}{l|cccc|ccc|cc|cc|cc}
        \toprule
        \multirow{3}{*}{\textbf{Model}}  & \multicolumn{7}{c|}{\textbf{Cascade}} & \multicolumn{6}{c}{\textbf{Direct}} \\
        \cmidrule{2-14}
        &\multicolumn{4}{c|}{\textbf{Level 1}}  &  \multicolumn{3}{c|}{\textbf{Level 2}} &  \multicolumn{2}{c|}{ \textbf{Level 1}} & \multicolumn{2}{c|}{ \textbf{Level 2}}&  \multicolumn{2}{c}{ \textbf{Level 3} }\\
        \cmidrule{2-14}
        &\textbf{CL} & \textbf{SK} & \textbf{Code} & \textbf{PV} & \textbf{SK} &  \textbf{Code}& \textbf{PV}  & \textbf{Code} & \textbf{PV} & \textbf{Code} & \textbf{PV} & \textbf{Code} & \textbf{PV} \\
        \midrule
        \multicolumn{14}{c}{\textbf{Open-source AGI LLMs}}\\
        \midrule
        Llama-2-7B & 1.13 & 0.43 & 0.32 & 0.00 & 0.00 & 0.49 & 0.00 & 0.13 & 0.00 & 0.30 & 0.00 & 0.40 & 0.00 \\ 
        Llama-3.1-7B & 3.61 & 2.60 & 1.41 & 0.33 & 1.92 & 1.99 & 0.68 & 1.92 & 0.68 & 1.84 & 0.39 & 4.74 & 0.54 \\ 
        Llama-3.2-3B & 1.00 & 0.27 & 1.61 & 0.00 & 0.55 & 0.00 & 0.00 & 0.17 & 0.00 & 0.00 & 0.00 & 0.17 & 0.00 \\ 
        Phi-3-14B & 5.62 & 1.69 & 1.37 & 0.44 & 1.14 & 0.88 & 0.00 & 0.30 & 0.00 & 0.00 & 0.00 & 0.29 & 0.00 \\ 
        Phi-4-14B & \textbf{41.92} & 3.74 & 1.71 & 1.37 & 2.42 & 3.82 & 4.74 & \textbf{10.87} & \textbf{10.76} & \textbf{6.87} & \textbf{8.15} & \textbf{13.32} & \textbf{9.70} \\ 
        Gemma-7B & 38.08 & 5.70 & 5.12 & 0.00 & 6.98 & 6.37 & 0.00 & 1.95 & 0.00 & 2.38 & 0.00 & 5.38 & 0.00 \\ 
        Gemma-2-9B & 40.25 & \textbf{7.90} & 7.53 & 9.32 & 7.70 & \textbf{9.05} & 6.62 & 5.50 & 8.93 & 5.97 & 6.59 & 8.07 & 5.19 \\ 
        Mistral-7B-v0.3 & 4.20 & 7.12 & \textbf{8.73} & \textbf{9.74} & \textbf{7.03} & 7.48 & \textbf{7.14} & 6.37 & 6.16 & 6.80 & 7.45 & 7.81 & 7.23 \\ 
        \midrule
        \multicolumn{14}{c}{\textbf{Code Generation LLMs}}\\
        \midrule
        StarCoder-2-7B & - & - & - & - & - & - & - & - & - & - & - & 0.00 & 0.00 \\ 
        CodeGemma & - & - & - & - & - & - & - & - & - & - & - & \textbf{9.99} & \textbf{14.75} \\ 
        CodeLlama & - & - & - & - & - & - & - & - & - & - & - & 5.44 & 0.99 \\ 
        \midrule
        \multicolumn{14}{c}{\textbf{Production Code Agent}}\\
        \midrule
        OpenHands*\dag & - & - & - & - & - & - & - & 22.96 & 35.51 & - & - & - & - \\ 
        \midrule
        \multicolumn{14}{c}{\textbf{Close-source AGI LLMs}}\\
        \midrule
        GPT-3.5-turbo  & 38.33 & 8.82 & 13.73 & 38.46 & 12.56 & 13.55 & 42.30 & 13.27 & 37.73 & 13.91 & 41.19 & 34.19 & 39.21 \\ 
        GPT-4o & \textbf{\underline{55.73}} & \textbf{\underline{16.57}} & \textbf{\underline{36.37}} & \textbf{\underline{54.75}} & \textbf{\underline{15.46}} & \textbf{\underline{36.42}} & \textbf{\underline{53.62}} & \textbf{\underline{35.18}} & \textbf{\underline{51.75}} & \textbf{\underline{33.10}} & \textbf{\underline{50.16}} & \textbf{\underline{53.01}} & \textbf{\underline{62.69}} \\ 
        Gemini 1.5 pro  & 49.48 & 14.01 & 31.96 & 18.15 & 15.22 & 31.04 & 25.62 & 15.97 & 9.99 & 24.32 & 22.05 & 46.51 & 27.90 \\ 
        Gemini 2.0-flash & 51.85 & 16.08 & 20.63 & 6.61 & 17.63 & 22.02 & 11.53 & 26.39 & 10.46 & 24.99 & 13.39 & 41.89 & 19.40 \\ 
        Gemini 2.0-pro* & 49.44 & 13.69 & 16.86 & 5.09 & 19.93 & 24.09 & 10.91 & 2.61 & 0.00 & 30.19 & 16.22 & 36.95 & 10.39 \\  
        \midrule
        \centering Average & 29.28 & 7.59 & 11.34 & 11.10 & 8.35 & 12.09 & 12.55 & 9.28 & 10.50 & 11.59 & 12.74 & 16.76 & 12.37\\
        \bottomrule
        \end{tabular}
    }
    \caption{ProjectEval Result Objective Indicators. Phi-4, the Gemmas and all close-source LLM agents have abilities to generate Checklist well, but only close-source LLM agents can do the Skeleton and Code well. CL: Checklist, SK: Skeleton, PV: Parameter Value. * We only test the Gemini-2.0-pro and OpenHands pass@1. \dag We used GPT-4o as Openhands' based LLM.}
    \label{tab:objective-result}
\end{table*}
\subsection{RQ1: Overall Correctness}
\label{overall-correctness}

Table~\ref{tab:pass@5} shows the overall correctness is low. The results are very similar to CoderEval\citeyearpar{yu2024codereval} and Dev-Bench\citeyearpar{li2024devbench} as all agents are very unlikely to make the project runnable (only 17.91\% projects in CoderEval’s result) and almost impossible to make every details correct in the project (lower than 10\% passed in DevBench's result). 

Table~\ref{tab:objective-result} shows that most of the opensource-model agents cannot generate compilable project. We examine through Phi-4 and Gemma-2, they have only 1 to 3 simple projects can be compiled and run. Even if the close-source LLMs can hardly reach the 10\% of ProjectEval standard, but the close-source LLM agents do better than the open-source ones which is equivalent to 1 or 2 simple projects get almost full scores.

However, Table~\ref{tab:objective-result} shows that the Gemma, Gemma-2 and Phi-4 do have the abilities on Checklist generating as they have approximate score to GPT-3.5-turbo, lower than GPT-4o and Geminis, while the Llama series, Mistral and Phi-3 have very low scores. The latter is caused by the lack of JSON format adaptability.

Both tables shows that Gemma-2 and Phi-4 may have the same capability to GPT-3.5-turbo, but far more way to go for GPT-4o and Gemini-1.5-pro.

We examine the logs of the LLMs and they shows that the typical failure cases are:
\begin{itemize}
    \item \textbf{Invalid Output Format}: LLMs sometimes produce outputs that do not conform to the expected format. This is more common with the Llama series, potentially due to a lack of training data on formatted code and JSON.
    \item \textbf{Missing Essential Files}: Essential files, such as manage.py in Django projects, are often omitted. This may be because these files are considered basic and are therefore underrepresented in training corpora which may more be focusing on function files and their problem.
    \item \textbf{Omitted Content}: LLMs frequently leave sections blank or only comments. This may be due to the prevalence of community corpora where answerers often leave custom parts blank for the questioner to fill in.
    \end{itemize}

Additionally, the Code LLMs agents have almost no effective results can be produced. The reason may be that Skeleton only has natural language descriptions, and it is difficult to fill in the whole framework without the context code.

We also test the current SOTA production coding agent, the OpenHands \citep{wang2025openhands}, by using GPT-4o as its based LLM. It shows that there were 8 tasks that OpenHands agent did not finish (Task 2, 8, 10, 11, 14, 20 of AgentStuckInLoopError and Task 16, 17 of Not Complete) and lacks a lot of optional features and pages. As ProjectEval's tasks are origin and have no user interaction while the agents generate the code, it is acceptable that OpenHands got lower scores than the source GPT-4o (See Appendix \ref{sec:pass@k}), but it still beat the Gemini-1.5-pro which indicates that coding agents may have a better chance to pass ProjectEval.

In summary, ProjectEval is hard for nowadays agents as only GPT-4o reach the Pass@5 of 15\%. Open-source LLM agents are doing worse than the close-source ones, and Code LLMs agents do not have the abilities to pass ProjectEval.

\subsection{RQ2: Cascade Generation \& Direct Generation}

The average scores of objective indicators in Table~\ref{tab:objective-result} show that agents are doing better on cascade generation than the direct generation with 2.06\% higher at Level 1 input. The cascade generation in a way mimics the CoT process and the ReAct design of an agent. It allows agents to re-examine the project development procedure and correct some errors. As for the core scores (Pass@K) of all LLMs, they are too low to analyze. But we notice that the Gemmas, Mistral, and Gemini-1.5-pro are doing better in cascade generation. 

However, GPT-4o has higher scores when using direct generation mode rather than the cascade mode. So, we study a case, Project 3 – “Create a password generator”, of GPT-4o (See Appendix~\ref{sec:casestudy}). We resend all the input and output by order back to GPT-4o and ask the CoT of it. It shows that GPT-4o directly hits the files that need to be generated when using the direct generation mode while it concentrates more on the NL processing and analysis on cascade generation mode. This is an interesting phenomenon, and we suspect that asking the LLM agents to generate according to the thought steps we set induces the LLM to tend to activate parameters about natural language rather than the more important aspects of code generation.  

This phenomenon does not affect ProjectEval's evaluating capabilities as the cascade mode is not for ProjectEval core functions as the standard ProjectEval score is evaluating the projects generated by asking the LLM only once.

Also, this finding in ProjectEval Pass@5 is conflicted with the CodeBLEU result, as the latter’s cascade scores are higher than the direct scores no matter it uses Level 1 or Level 2 input, both in Code and PV. This means even if CodeBLEU has considered the structure, “details will determine success or failure”. For instance, we found that the GPT-4o's cascade generation did has better structure of the code but it just missed filling a path parameter and the result was fatal for the project.

In summary, cascade generation is better than direct generation, and ProjectEval execution pass rate is better than the similarity indicators as the latter cannot reflect the program execution effects. 

\subsection{RQ3: Basic LLM Selection}
The close-source LLM agents have better performance on ProjectEval than the open-source ones (See Table~\ref{tab:pass@5}). GPT-4o are the SOTA of project generation under ProjectEval evaluation. 

The first difference is the ability to generate systematic project code based on natural language. From the objective indicators (Table~\ref{tab:objective-result}), we find the close-source LLM agents do better on skeleton and code generation, both GPT and Gemini can generate better skeleton reflecting well to the standard code. We used GPT-4o as the ProjectEval's data generation procedure base model but Gemini-1.5-pro reaches almost the same performance of GPT-4o. Thus, the reason may not be the familiarity of GPT’s prompt. This suggests that close-source LLM agents may have better ability on reflecting the Checklist into a skeleton or framework. The Checklist and Skeleton’s Functions and/or Classes are many-to-many relationship. It is a very complicate mission for open-source LLM agents to solve.

Second, the over-all understanding of the project and comprehensive analysis capabilities are also very important for the LLM agents. Close-source LLM agents are doing well on all parts of the ProjectEval inputs and the open-source LLM agents will have a better chance to get passed ProjectEval when they have better performance on those parts.

Third, already mentioned in Section~\ref{overall-correctness},  the LLM agents’ formatted output capabilities have huge influence. We ask all LLM agents to output JSON format. All the close-source agents can do well compare with only Phi-4, Gemma and Gemma-2 have this capabilities of open-source agents. This finding is important for research or engineering that requires the LLM Agent to be the controller, as they need to guarantee stable and regulated outputs.

In summary, the close-source LLM agents are doing better on ProjectEval. GPT-4o are the SOTA of project generation under ProjectEval evaluation.

\subsection{RQ4: Step-by-Step Performance}
Objective indicators can also show the step-by-step performance in cascade generation (See Table~\ref{tab:objective-result}). 

Except the Llama series, from the cascade perspective, the LLM agents perform well in the field of NL generation for Checklists, which is also one of its fundamental capabilities. In addition, the Checklist itself includes an understanding of project prompts which means most LLMs agents also possess this capability. 

When it comes to the Skeleton, we have already mentioned the problem of many-to-many relationship question of Checklist and Skeleton. This is a challenge for LLM agents to deal with. And Code generation is a traditional topic of LLM agents, Phi-4 does better when it directly generates the code from Level 1 rather than the cascade. 

Parameter answering reflects the code understanding capability. Though, in ProjectEval, the score may highly connect with the Code generations but still may indicate that the capability of LLM agents. When the code is effective, both open-source and close-source LLM agents can explain their own code, which confirms that LLM agents have a strong ability to understand code. Specifically, it can identify the key statements needed from the PD as for ProjectEval questions.

In summary, LLM agents are best at generating Checklists than the other parts of ProjectEval.

\section{Conclusion}\label{sec:conclusion}
We develop a new benchmark ProjectEval. It fills the gap for the lack of benchmark in the project granularity level code generation field of natural language processing and provides automated evaluation tools for higher-level research in LLM agent. We also leverage additional objective metrics to reveal the effectiveness of LLM agents at different stages of project generation. These metrics are crucial in revealing the capabilities for improvement in the agents' performance, thus offering a deeper understanding of how these models can be further enhanced. We confirmed that GPT-4o is still the SOTA in this field. Our findings and benchmark provide valuable insights for developing more effective programming agents that can be deployed in real-world production environments. 

\newpage
\section{Limitation}\label{sec:limitation}
\begin{itemize}
    \item Some projects, whether due to complexity or human-related difficulties, may follow common design patterns that do not align with Django, our primary technical stack. 
    \item The canonical answer right now only supports Python language but in theory, any language that can be rendered and simulated can be tested by the Test suite.
    \item JSON format is not universally compatible with all LLMs. 
    Some models might perform better if they were allowed to generate output in their own format. However, permitting this would compromise fairness and introduce inconsistencies in output standards. 
    \item CPV is based on the CNC, but the CNC may not the only answer for the project. This may lead to some PV is correct for the reflected code but will get lower score on Levenshtein Distance evaluation. 
\end{itemize}
All these limitations will be solved in the future research if it is possible.

Besides, the ProjectEval judge machine will automatically run the project that generated by LLM agents which may contain harmful code. This is a potential risk that we can't fix. There are also some issues that this work does not fully capture like software maintainability, efficiency, or best coding practices.
\section*{Acknowledgments}
This work is supported by the National Science and Technology Innovation 2030 Major program  (Grant No. 2024ZD01NL00101).

\bibliography{custom}

\appendix

\section{ProjectEval Structure and Construction Process Full-version}
\begin{figure*}[ht]
    \centering
    \includegraphics[width=1\linewidth]{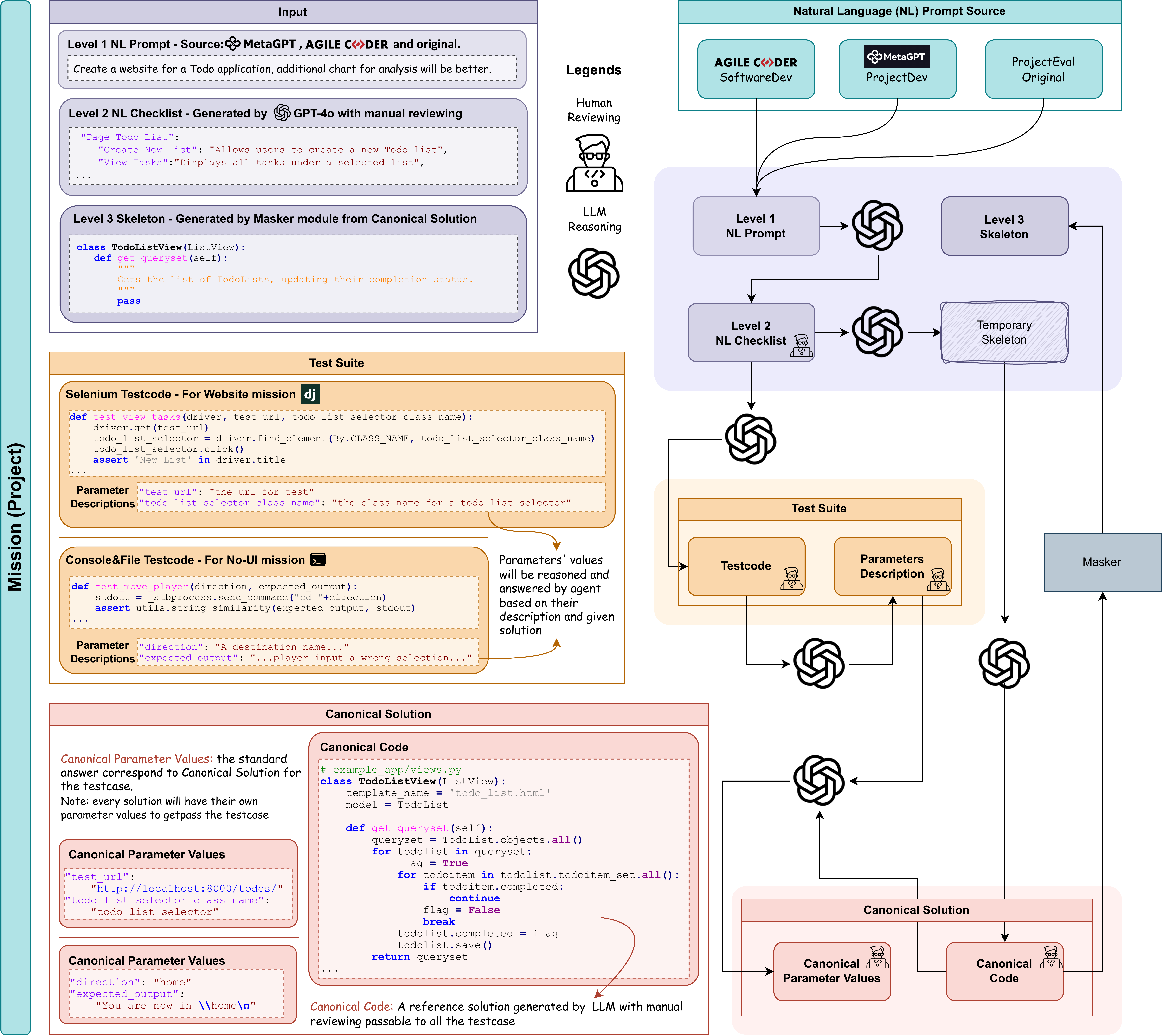}
    \caption{ProjectEval Structure and Construction Process}
    \label{fig:full-figure}
\end{figure*}

Figure~\ref{fig:full-figure} shows the full-version of the structure and construction process of ProjectEval.

\label{sec:full-figure}

\section{Basic LLM Selection and Settings}
\label{sec:selection}
Since the ProjectEval Level 1, Level 2 input and PV input are natural language but the Level 3 input, and code output are program language, the basic LLM models should have both aspects abilities. This means that ProjectEval full evaluation must be run under AGI LLM models. 

We select three types of models: open-source AGI LLMs, close-source AGI LLMs, and code generation LLMs. Among the open-source AGI models, we include Mistral-7B-v0.3\citeyearpar{mistral}, Gemma\citeyearpar{gemma}, Phi\citeyearpar{phi}, and Llama\citeyearpar{llama}, which are known for their advancements in general AI capabilities. In the close-source AGI LLM category, we consider models such as GPT\citep{openai} and Gemini\citep{google}, which represent cutting-edge proprietary models excelling in a variety of tasks. 

Additionally, the Code generation LLMs category features CodeLlama and Starcoder2, which are specialized in code generation and will be used only in Skeleton input evaluation. 

For Gemma series, we include Gemma-7B, Gemma2-9B; For Phi series, we include Phi-4, Phi-3-14B; For Llama series we include Llama3.2-3B, Llama3.1-8B, Llama2-7B; For GPT series, we include GPT-4o and GPT-3.5-turbo; For Gemini we include Gemini-1.5-pro and Gemini-2.0. All the models are running under temperature zero with all settings default in their releases.

\section{Related Benchmarks}
\label{sec:benchmarks-fullversion}
\begin{table*}[ht!]
\centering
\renewcommand\arraystretch{1.5}
\setlength{\tabcolsep}{2pt}
\tiny
\begin{tabular}{@{}c|cccm{2cm}<{\centering}m{1.2cm}<{\centering}ccccm{2cm}<{\centering}@{}}
\toprule
\textbf{Benchmark} & \textbf{Language} & \textbf{Construction} & \textbf{Evaluation} & \textbf{Source} & \textbf{Granularity Level} & \textbf{\#Tasks} & \textbf{\#Tests} & \textbf{\#LOC} & \textbf{\#Tokens} & \textbf{Input Information} \\
\midrule
\multicolumn{11}{c}{Early Years Research} \\
\midrule
Concode \citeyearpar{iyer2018concode} & Java & Automated & Automated & GitHub & Function & 2000 & N/A \footnotemark[1] & N/A \footnotemark[1] & 26.3 & NL \\
CoNaLA \citeyearpar{yin2018conala} & Python & Automated & Automated & Stack Overflow & Statement & 500 & N/A \footnotemark[1] & 1.0 & 4.6 & NL \\
\midrule
\multicolumn{11}{c}{BLEU-based}\\
\midrule
Django \citeyearpar{iyer2018django} & Python & \multirow{3}{*}{N/A\footnotemark[2]} & \multirow{3}{*}{Automated} &\multirow{3}{*}{N/A\footnotemark[2]} & \multirow{3}{*}{Project} & \multicolumn{4}{c}{\multirow{3}{*}{N/A\footnotemark[2]}} & \multirow{3}{*}{Code} \\
CodeBLEU\citeyearpar{ren2020codebleu} & Multiple & &  & & & &  &  & & \\
SketchBLEU\citeyearpar{zan2024codes} & Multiple & &  & & &  &  & & & \\
\midrule
\multicolumn{11}{c}{HumanEval-based}\\
\midrule
HumanEval \citeyearpar{chen2021humaneval} & Python & Manual & Automated & Original & Function & 164 & 7.7 & 11.5 & 24.4 & NL+ Function Signature \\
AixBench \citeyearpar{hao2022aixbench} & Java & Manual & Automated & HumanEval \& Original & Function & 175 & \multicolumn{3}{c}{N/A\footnotemark[3]} & NL+ Function Signature \\
Multi-HumanEval\citeyearpar{athiwaratkun2023multihumaneval} & Multiple & Manual & Automated & HumanEval \& Original & Function & 164 & 7.7 & 11.5 & 24.4 & NL+ Signature \\
HumanEval+ & Python & Manual & Automated & Original & Function & 164 & 774.8 & 11.5 & 24.4 & NL+ Signature \\
\midrule
\multicolumn{11}{c}{MBPP-based}\\
\midrule
MBPP\citeyearpar{austin2021mbpp} & Python & Manual & Automated & Original & Function & 974 & 3.0 & 6.8 & 24.2 & NL \\
APPS\citeyearpar{hendrycks2021measuring} & Python & Automated & Automated & Contest Sites & Competitive & 5000 &13.2 & 21.4 &58 & NL+ Examples \\
MBXP\citeyearpar{athiwaratkun2023multihumaneval} & Multiple & Manual & Automated & MBPP \& Original & Function & 974 & 3.0 & 6.8 & 24.2 & NL \\
CodeContests\citeyearpar{li2022codecontests} & Python, C++ & Automated & Automated & Contest Sites & Competitive & 165 & 203.7 & 59.8 & 184.8 & NL + Example I/O \\
\midrule
\multicolumn{11}{c}{DS-based}\\
\midrule
DS-1000\citeyearpar{lai2023ds1000} & Python & Automated & Automated & Stack Overflow & Statement & 1000 & 1.6 & 3.8 & 12.8 & NL \\
CoderEval\citeyearpar{yu2024codereval} & Python, Java & Automated & Automated & Github & Function & 230 & N/A\footnotemark[1] & 30.0 & 108.2 & NL + Function Signature \\
ClassEval \citeyearpar{du2023classeval}& Python & Manual & Automated & PyPI + Original & Class & 100 & 33.1 & 45.7 & 123.7 & Class Skeleton \\
EvoCodeBench \citeyearpar{li2024evocodebench}& Python & Semi-automated & Automated & GitHub & Function & 275 & N/A \footnotemark[1]  & 20.40 &  185.57&  Repository\\
SWEBench\citeyearpar{jimenez2024swebench} & Python & Automated & Automated & PyPI & Commit & 2294 & 120.8 & 32.8 & $\approx$ 200 & NL + Repository \\
Commit0\citeyearpar{zhao2024commit0} & Python & Automated & Automated & PyPI & Commit & 54 & $\approx$ 500 & $\approx$ 25000 & $\approx$ 1250 & NL + Testunits
\\
\midrule
\multicolumn{11}{c}{Program Understanding}\\
\midrule
ReCode\citeyearpar{wang-etal-2023-recode} & Python & Automated & Automated & HumanEval \& MBPP & Function& 30 & 10.0 & N/A \footnotemark[1]  & N/A \footnotemark[1]  & Code \\
CRUXEval \citeyearpar{gu2024cruxeval} & Python & Automated & Automated &  Python Standard Libs & Function & 800 & 10.0 & 5.49 &  N/A \footnotemark[1] & Code \\
CodeBenchGen\citeyearpar{xie2024codebenchgen}& Python & Automated & Automated & CodeSearchNet\citeyearpar{husain2019codesearchnet} (GitHub) & Function & 1931 & 8.79 & 60.5 & 491.9 & Code + NL Statements\\ 
\midrule
\multicolumn{11}{c}{Granularity Level - Project}\\
\midrule
SRDD \citeyearpar{qian2024srdd}& Python & Automated & Automated & Original & Project & 1200 & N/A \footnotemark[1] & N/A \footnotemark[1] & N/A \footnotemark[1] & NL \\
CASSD \citeyearpar{zhang2024cassd} & Python & Manual & Manual& Original & Project & 72 & 5.25 & $\approx$240 & 21993.0 & NL \\
SoftwareDev\citeyearpar{hong2024metagpt} & Multiple & Manual\footnotemark[1] & N/A & Original & Project & 70 & N/A \footnotemark[1] & 191.6 & 6218.0 & NL \\
ProjectDev\citeyearpar{nguyen2024agilecoder}  & Multiple & Manual & Manual & Original & Project & 14 & 19.1 & N/A \footnotemark[1] & 36818.0 & NL \\ 
DevBench\citeyearpar{li2024devbench} & Multiple & Manual & Automated & Original & Project & 22 & 10.18 & 377.8 & 1298.3 & NL \\
\textbf{ProjectEval (ours)} & \textbf{Python} & \textbf{Semi-automated} & \textbf{Automated} & \textbf{SoftwareDev \& ProjectDev \& Origin} & \textbf{Project} & \textbf{20} & \textbf{14.2} & 402.2 & 2972.0 & \textbf{NL+ Class/Function Skeleton} \\
\bottomrule
\multicolumn{11}{l}{[1]This benchmark doesn’t involve this item.}\\
\multicolumn{11}{l}{[2]The Bleu-based benchmark doesn’t involve construction, source and number items.}\\
\multicolumn{11}{l}{[3]Since Aixbench wrote 175 Java files for testing and wrote the test samples directly into the code, it is very difficult to count its details.}\\
\end{tabular}
\caption{Summary of Existing Benchmarks for Code Generation. \#Tasks: number of tasks, \#Tests: average number of testcase in each task, \#LOC: average lines of code in the canonical answer, \#Tokens: average number of tokens of code in the canonical answer. Part of this table is referred from \citeposs{du2023classeval}.}
\label{tab:benchmarks-fullversion}
\end{table*}
Table~\ref{tab:benchmarks-fullversion} is the full-version of the related benchmarks.

HumanEval-based Benchmarks: These benchmarks are similar to or extensions of OpenAI's HumanEval, primarily emphasizing functional tasks with general-purpose functions.

MBPP-based Benchmarks: These benchmarks are similar to or derived from Google's MBPP, focusing on algorithmic problem-solving functions.

DS-based Benchmarks: DS-based benchmarks involve the use of external libraries or classes. However, the specific contents and documentation of these libraries/classes are not included within the benchmark dataset itself.

Program Understanding Benchmarks: Unlike other benchmarks that focus on whether the code is written correctly, these benchmarks assess the ability of an agent or LLM to thoroughly understand the provided code.

Project-level Benchmarks: Project-level benchmarks, which are the primary focus of this paper, comprehensively evaluate the process of transforming an initial idea into a complete program. There are currently five benchmarks in this category, see Table~\ref{tab:compare} for the differences. 

\section{Prompts \& Outputs}
\label{outputs}
Here is the prompt of the Reasoning:

\begin{tcolorbox}[colback=cyan!10,colframe=cyan!40!black, title=\textbf{ProjectEval Reasoning Prompt}]
Based on this \{description\}, give a \{technical\_stack\} Project of its all files (including the essential files to run the project) to meet the requirement in JSON format of \texttt{[ \{"file":"answer.something",\allowbreak "path":\allowbreak "somepath/somedir/answer.something", "code":"the\_code\_in\_the\_file"\},\{...\},\allowbreak ... ]} with \textbf{NO other content}. Recommend adding an \texttt{id} attribute to each HTML element and adding classes for them too.
\end{tcolorbox}

The reasoning phase will eventually come to an JSON file. Here is an example of the output:
\begin{tcolorbox}[title={An Example of ProjectEval Output},colback=olive!10,colframe=olive!70!black]
\begin{lstlisting}[breaklines=true, columns=fixed, flexiblecolumns,basicstyle=\small\ttfamily]
{
    "19": [
        {
            "file": "19-stdanswer.py",
            "path": "19-stdanswer.py",
            "code": "import pandas as pd\n..."
        }
    ]
}
\end{lstlisting}
\end{tcolorbox}

\section{Human Annotators Instruction}
\label{human-instruction}
\begin{figure*}[ht]
    \centering
    \includegraphics[width=1\linewidth]{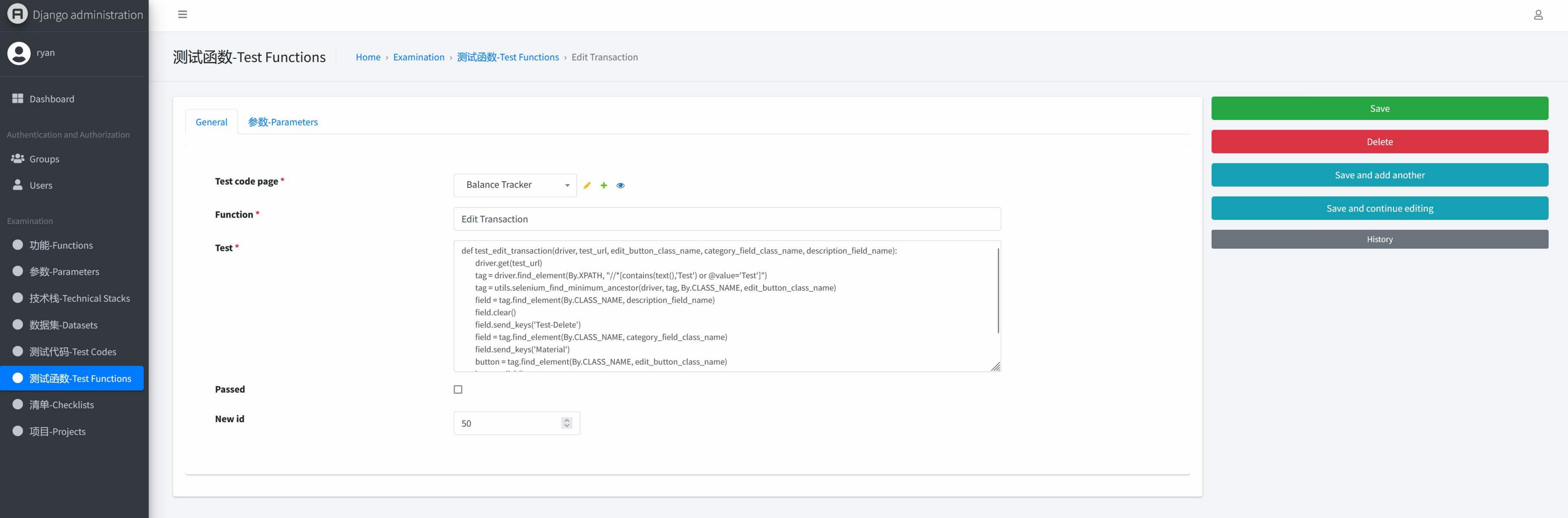}
    \caption{screenshot of the Assisting Websystem}
    \label{fig:screenshot}
\end{figure*}

Our human annotators have an 8-page instruction guideline for them to annotate the code in both the Canonical Solutions and the Test Suites. It is impossible to show it in the paper and in the appendix as it is in Chinese and it is too long with instruction images and an assisting small information web system. As compensation, we provide the following annotation key steps (listed in the instruction guideline):
\begin{enumerate}
    \item Review the checklist, see if there are any inappropriate requirements and remove them or edit them, and add the missing general requirements;
    \item Review the answer code in render mode, and make sure it can meet all the requirements of the annotated checklist;
    \item Run the answer code and review the test code and the parameter values in render mode, make sure it can test the checklist requirements, and those common-sense requirements (such as login authorization check), since the common-sense requirements may not be added in Step 2, you may edit the answer code at the same time;
    \item Review the parameter description, add the missing ones, and remove the redundant ones;
    \item Run the entire answer code again, make sure that all the tests can be aligned with the answer.
\end{enumerate}
Figure~\ref{fig:screenshot} shows a screenshot of the assisting small information web system.

\section{ProjectEval Result Pass@K}
Table~\ref{tab:pass@k} is the full result of the ProjectEval.
\label{sec:pass@k}
\begin{table*}[hp!]
\centering
    \renewcommand{\arraystretch}{1.5}
    \small
    \setlength{\tabcolsep}{2pt}
    \resizebox{\textwidth}{!}{
    \begin{tabular}{l|cc|ccc|ccc|cccc|c}
        \toprule
        \multirow{3}{*}{\textbf{Model}}  & \multicolumn{5}{c|}{\textbf{Pass@1}} & \multicolumn{8}{c}{\textbf{Pass@5}} \\
        \cmidrule{2-14}
        &\multicolumn{2}{c|}{\textbf{Cascade}}  &  \multicolumn{3}{c|}{\textbf{Direct}} &  \multicolumn{3}{c|}{ \textbf{Cascade}} & \multicolumn{4}{c|}{ \textbf{Direct}}&  \multirow{2}{*}{ \textbf{All Avg.} }\\
        \cmidrule{2-13}
        &\textbf{Level 1} & \textbf{Level 2} & \textbf{Level 1} & \textbf{Level 2} & \textbf{Level 3} &\textbf{Level 1} & \textbf{Level 2}  & \textbf{Avg.} & \textbf{Level 1} & \textbf{Level 2} & \textbf{Level 3} & \textbf{Avg.} &  \\
        \midrule
        \multicolumn{14}{c}{\textbf{Open-source AGI LLMs}}\\
        \midrule
        Llama-2-7B & 0.00 & 0.35 & 1.06 & 0.00 & 0.35 & 0.00 & 0.07 & 0.04 & 0.28 & 0.00 & 0.07 & 0.12 & 0.08 \\ 
        Llama-3.1-7B & 0.70 & 0.70 & 0.35 & 0.70 & 1.06 & 0.28 & 0.28 & 0.28 & 0.14 & 0.28 & 0.42 & 0.28 & 0.28 \\ 
        Llama-3.2-3B & 0.70 & 0.35 & 0.35 & 0.00 & 0.00 & 0.21 & 0.14 & 0.18 & 0.14 & 0.00 & 0.00 & 0.05 & 0.10 \\ 
        Phi-3-14B & 0.00 & 0.00 & 0.00 & 0.00 & 0.00 & 0.00 & 0.00 & 0.00 & 0.00 & 0.00 & 0.00 & 0.00 & 0.00 \\ 
        Phi-4-14B & 0.35 & 1.06 & \textbf{4.23} & \textbf{1.41} & \textbf{4.23} & 0.14 & 0.56 & 0.35 & \textbf{1.76} & \textbf{1.13} & \textbf{2.04} & \textbf{1.64} & \textbf{1.13} \\ 
        Gemma-7B & 1.41 & 1.41 & 1.06 & 0.70 & 1.06 & 0.99 & 1.06 & 1.02 & 0.56 & 0.63 & 0.49 & 0.56 & 0.75 \\ 
        Gemma-2-9B & \textbf{2.11} & \textbf{2.46} & 2.11 & 1.06 & 1.06 & \textbf{1.69} & \textbf{1.06} & \textbf{1.37} & 1.34 & 0.56 & 0.63 & 0.85 & 1.06 \\ 
        Mistral-7B-v0.3 & 2.11 & 1.76 & 1.41 & 1.41 & 0.70 & 1.48 & 1.06 & 1.27 & 0.92 & 0.99 & 0.56 & 0.82 & 1.00 \\ 
        \midrule
        \multicolumn{14}{c}{\textbf{Code Generation LLMs}}\\
        \midrule
        StarCoder-2-7B & - & - & - & - & 0.00 & - & - & - & - & - & 0.00 & - & - \\ 
        CodeGemma & - & - & - & - & \textbf{2.11} & - & - & - & - & - & \textbf{1.20} & - & - \\ 
        CodeLlama & - & - & - & - & 1.41 & - & - & - & - & - & 0.77 & - & - \\ 
        \midrule
        \multicolumn{14}{c}{\textbf{Production Code Agent}}\\
        \midrule
        OpenHands & - & - & 7.39 & -  & -  & -  & - & - & - & - & - & - & - \\
        \midrule
        \multicolumn{14}{c}{\textbf{Close-source AGI LLMs}}\\
        \midrule
        GPT-3.5-turbo  & 2.46 & 2.46 & 2.46 & 2.46 & 6.69 & 2.39 & 2.46 & 2.43 & 1.97 & 2.39 & 5.28 & 3.22 & 2.90 \\ 
        GPT-4o & \textbf{\underline{10.21}} & \textbf{\underline{15.85}} & \textbf{\underline{19.72}} & \textbf{\underline{17.96}} & \textbf{\underline{12.32}} & \textbf{\underline{8.52}} & \textbf{\underline{12.32}} & \textbf{\underline{10.42}} & \textbf{\underline{16.06}} & \textbf{\underline{15.42}} & \textbf{\underline{10.14}} & \textbf{\underline{13.87}} & \textbf{\underline{12.49}} \\ 
        Gemini 1.5 pro  & 9.15 & 8.80 & 7.39 & 6.34 & 9.51 & 7.82 & 7.39 & 7.61 & 5.28 & 4.51 & 8.24 & 6.01 & 6.65 \\ 
        Gemini 2.0-flash & 5.63 & 5.28 & 5.63 & 6.69 & 8.80 & 3.24 & 3.59 & 3.42 & 3.52 & 3.45 & 7.75 & 4.91 & 4.31 \\ 
        Gemini 2.0-pro* & 4.93 & 4.93 & 0.00 & 5.63 & 7.39 & - & - & - & - & - & - & - & - \\ 
        \midrule
        Avg. & 3.06 & 3.49 & 3.52 & 4.03 & 3.54 & 2.23 & 2.50 & 2.37 & 2.66 & 2.45 & 2.51 & 2.69 & 2.56 \\ 
        \toprule
        \multicolumn{14}{l}{* We only test the Gemini-2.0-pro pass@1.} \\
    \end{tabular}
    }
    \caption{ProjectEval Result Pass@k}    
    \label{tab:pass@k}
\end{table*}

\clearpage
\onecolumn
\section{Casestudy Log of GPT-4o}
\label{sec:casestudy}
\subsection{Cascade Version}
\begin{lstlisting}[breaklines=true, basicstyle=\small\ttfamily, frame=single, caption={Cascade Verison},columns=fixed, flexiblecolumns]
My chain of thought for each step in the log is as follows:

1. **Instruction to Generate Natural Language Function Checklist**:
   - **Input Reasoning**: A request is made to create a website for a password generator and generate a natural language function checklist from the users' perspectives in a JSON format.
   - **Action Taken**: Based on the request, I generate a JSON object representing different pages of the proposed website, each containing functions for user interactions, along with descriptions of these functions.
   - **Output Reasoning**: I structure the output JSON to clearly separate the different pages (e.g., Home, Password Generator, Settings) and the functions associated with each, providing a concise description for each user-facing action.

2. **Generate Django Framework Skeleton**:
   - **Input Reasoning**: Build a Django framework skeleton based on the previously provided checklist, using JSON format to outline the necessary files and logic.
   - **Action Taken**: Construct a basic Django app structure, including `views.py` and `urls.py` for simplicity, containing placeholder functions that align with the provided natural language functions.
   - **Output Reasoning**: The JSON output details files like `views.py` and `urls.py`, including paths and basic skeleton code for each function as derived from the JSON function checklist.

3. **Complete Django Project with Essential Files**:
   - **Input Reasoning**: Expand the Django project to include all essential files for a basic working project, add ID attributes, and consider adding classes to HTML elements for future CSS or JavaScript interactions.
   - **Action Taken**: Create a comprehensive JSON representation of a Django project, including files such as `manage.py`, `settings.py`, `urls.py` for the project and the app, HTML templates, and CSS files necessary for the application.
   - **Output Reasoning**: The output JSON contains each file's relative path and code content, ensuring each crucial component for running a Django application is included. It also adheres to the instruction of adding identifiers like IDs and classes to HTML elements.

4. **Generating Required Parameters for Django Project Test**:
   - **Input Reasoning**: Derive and return necessary parameters for testing each function within the previously defined Django project based on the requirements outlined, using JSON format, specifying example URLs and identifiers.
   - **Action Taken**: I extract specific identifiers (like IDs and URLs) used in the Django project HTML and map them to their respective functions in a JSON format that defines parameter names and their corresponding examples.
   - **Output Reasoning**: The JSON output specifies each function's parameters for testing purposes, using IDs defined in the HTML and expected URL paths. Each function in the project maps back to a user scenario described earlier, ensuring the testing parameters align with those scenarios.

Throughout each step, I maintained a structured approach to interpret the instructions, generating JSON outputs that capture the interaction logic, application skeleton, and testing parameters in alignment with the requirements provided.
\end{lstlisting}
\subsection{Direct Version}
\begin{lstlisting}[breaklines=true, basicstyle=\small\ttfamily, frame=single, caption={Direct Verison},columns=fixed, flexiblecolumns]
Given the provided log, here's how I processed the task requests step-by-step:

### Initial Request and Reasoning

1. **Request to Provide a Django Project:**

   - The request was to create a Django project for a password generator. The specific structure was requested in a JSON format that included all necessary files for the project.
   - The project was constructed with essential modules and files, typical of a basic Django setup: `manage.py`, `__init__.py`, `settings.py`, `urls.py`, `wsgi.py`, etc.
   - Inside the project folder, two main components were established: `password_generator` for project-level configurations and `generator` for the app dedicated to password generation logic.

2. **Project Configuration and Application Code:**

   - **`manage.py`**: Standard entry point for running Django commands.
   - **`settings.py`**: Configures the project's settings including installed apps (like `generator`), middleware, database configurations, allowed hosts, static files handling, etc.
   - **`urls.py`** at the project level defines the URL routes and maps them to respective views in the `generator` app.
   - The app `generator` has core logic hosted in its `views.py`, handling requests to generate passwords with different options (uppercase, numbers, symbols).

3. **HTML Templates for User Interaction:**

   - **`home.html`**: Presents users with options to generate a password, using forms with checkboxes and a submit button to include various character types.
   - **`password.html`**: Displays the generated password, allowing users to generate another if desired.
   - **`about.html`**: Provides basic project information.
   - IDs and classes were assigned to HTML elements per the requirement to enable easy referencing in styles and scripts.

### Parameterized Test Outputs

4. **Parameters Extraction for Test:**

   - A structured JSON was requested to determine specific parameter values within the project for automated testing purposes.
   - Each HTML element's IDs, button names, and page URLs were mapped to ensure correct access and functionality tests.

5. **Page: Home:**

   - It confirmed the main URL and associated ID for displaying the welcome message (`title`).
   - Navigation from the home page via specific button IDs to the password generator feature was detailed.

6. **Page: PasswordOptions:**

   - Confirmed the identification of form elements like length input and checkboxes for including character types (uppercase, numbers, symbols) during password generation.

7. **Page: GeneratedPassword:**

   - Ensured the display of generated passwords was accurately linked via the element ID (`password`).
   - Proposed testing for a "Copy to Clipboard" functionality (though the ID `copy` was assumed, this component may have been developed later).

8. **Page: About:**

   - Validated navigation to and from the 'About' page. Extracted the ID from the hyperlink (`about`) as well as a description ID, although it posed wrong by using `title` where expected a unique ID for description was needed.

9. **Errors & Oversights:**

   - Some mismatches were found in the parameter mappings, notably in assuming generic IDs like `length` to also be suitable for lowercase inclusion function tests or reusing `title` for `description_id`.
   - This indicates that further adjustments or clarification might be needed to align data bindings for consistent testing.

This reflection covers the derivation of each function response from the given initial requests and input, along with the pathway the project's elements were stored and rationalized for potential tests.
\end{lstlisting}
\end{document}